\documentstyle[prl,aps,floats,twocolumn,graphics]{revtex}

\author{J. Houdayer and O.~C. Martin}
\title{Ising Spin Glasses in a Magnetic Field}
\address{LPTMS, b\^at. 100, 
Universit\'e Paris-Sud, F--91405 Orsay, France.}
\date{\today}
\draft

\begin{document}
\sloppy

\maketitle
\begin{abstract}
Ground states of the three dimensional Edwards-Anderson spin glass are
computed in the presence of an external magnetic field. Our algorithm is
sufficiently powerful for us to treat systems with up to $600$ spins. We
perform a statistical analysis of how the ground state changes as the field
is increased, and reach the conclusion that the spin glass phase at zero
temperature does not survive in the presence of any finite field. This is in
agreement with the droplet model or scaling predictions, but in sharp
disagreement with the mean field picture. For comparison, we also
investigate a dilute mean field spin glass model where an Almeida-Thouless
line is present.
\end{abstract}


A long standing controversy in the field of spin glasses concerns the
stability of the spin glass phase when a magnetic field is applied. Mean
field approaches~\cite{MezardParisi87b} as well as a number of
Monte Carlo simulations~\cite{MarinariNaitza98} suggest that the spin glass
phase survives as long as the magnetic field is not too large; an
Almeida-Thouless (AT) line~\cite{AlmeidaThouless78} then separates this
phase from a paramagnetic phase at large fields. On the contrary, scaling
approaches \cite{McMillan86,BrayMoore86,FisherHuse88} predict that the spin
glass phase will not survive the application of any finite magnetic field;
the system becomes paramagnetic as soon as the external field is non-zero.
This controversy remains even though there have been a number of
experimental studies (see references in
\cite{BinderYoung86,FischerHertz91}); indeed, since it is difficult to reach
equilibrium in these systems, many of the experimental measurements may be
affected by out of equilibrium artifacts.

More generally, the mean field and scaling approaches differ fundamentally
in their view of the nature of large scale excitations. In mean field, the
spin glass phase is characterized by numerous nearly degenerate yet widely
separated valleys. The energies at the bottom of these valleys are expected
to be random~\cite{MezardParisi87b} as in Derrida's random energy
model~\cite{Derrida81}; also, the characteristic size of the energy gap
separating the two lowest energy valleys should be constant, {\it i.e.},
should not grow with the system size. In the scaling approach, however,
there are no such valleys; instead, system-size excitations should have
energies which grow as a power of their size.

The question of which approach is ``correct'' for three-dimensional spin
glasses remains largely open. Our purpose here is to test numerically the
stability of the spin glass phase to the application of an external magnetic
field at zero temperature ($T=0$). To do so, we were led to develop a new
and very effective algorithm for computing spin glass ground states. Then,
we apply this numerical tool to spin glass models in the presence of a
magnetic field. Our main focus is on the three dimensional Edwards-Anderson
(EA) model~\cite{EdwardsAnderson75} with Gaussian nearest neighbor
couplings. For comparison, we also study an appropriately chosen dilute mean
field spin glass model. Our conclusion is that in three dimensions the spin
glass phase does not survive the introduction of a finite magnetic field,
whereas it does survive in mean field models.

\paragraph*{Finding ground states ---}
We consider an EA Hamiltonian coupling nearest neighbor spins
on a three-dimensional cubic lattice with periodic boundary conditions:
\begin{equation}
H_J(\{S_i\}) = - \sum_{<ij>} J_{ij} S_i S_j - B \sum_{i} S_i.
\end{equation}
The first sum is over all bonds of the lattice, {\it i.e.}, over all
nearest neighbor spin pairs; $B$ is the external magnetic field, and the
$J_{ij}$'s are quenched couplings. We take these couplings to be independent
random variables with a Gaussian distribution of zero mean and unit
variance. The $J_{ij}$'s generate quenched disorder, and since the sign of
these couplings varies, they introduce frustration.

Finding the ground states of this system is {\it
NP}-hard~\cite{PapadimitriouSteiglitz82}. Not surprisingly then, even the
state of the art algorithms~\cite{SimoneDiehl96} for finding ground states
are very slow and are not able to effectively go beyond $5\times 5 \times 5$
lattices. Because of this, studies of spin glass ground states have
often~\cite{Hartmann98} used ``heuristic'' algorithms which only find the
ground state with a high probability. For our study, we have developed a
new heuristic algorithm which is based on iteratively optimizing the spin
configurations on {\it all} scales using a recursive genetic algorithm. It
is not guaranteed to find the lowest state, but we have performed checks to
measure its power and reliability~\cite{HoudayerMartin99}. In particular, we
have used the magnetic field to derive a self-consistency check as follows.
We start with a large magnetic field and compute the putative ground state.
This ground state is then used as a starting point for our iterative
improvement algorithm at a lower value of the field. We can then check for
consistency by going to negative values of the field and seeing whether the
states found are exactly the same (up to a global flip of the spins) as
those obtained at positive values of the field. If they are not, we
reinitiate our genetic algorithm with a larger population. Using this
approach, we estimate the error rate for finding the true ground state to be
less than 1 in $10^5$.

\paragraph*{Effect of the field ---}
For each lattice size and realization of the disorder variables $J_{ij}$'s,
we have determined the ground state configuration as a function of the
magnetic field $B$. For large $B$, nearly all spins are parallel to the
field. As the field is decreased continuously, the ground state changes at a
series of values of $B$; these values can be thought of as comprising a
discrete and finite ``spectrum'' for each instance. For a given point of
this spectrum, let $s$ be the number (or volume) of spins which are flipped
when going from one ground state to the next. Since the $J_{ij}$'s are
continuous, these $s$ spins form a connected cluster with probability $1$.
(Note that the ground state is generically non degenerate as soon as $B \ne
0$.) We focus our study on the statistical properties of $s$ as a function
of $B$. We have also measured the total magnetization associated with these
clusters of $s$ spins.

Before giving our results, let us describe qualitatively what is predicted
to occur in the scaling and mean field pictures. Let us begin with $B=0$. In
the scaling picture, it is useful to introduce the notion of a
droplet~\cite{FisherHuse88}. For each site and each length $l$, consider the
excitation of volume $O(l^3)$ which has the lowest energy and contains the
spin at that site. This excitation corresponds to flipping a connected
cluster of spins and is called a droplet. In the scaling approach, hereafter
called the {\it droplet picture}, the energy of a droplet scales as
$l^{\theta}$ as its ``diameter'' $l$ grows.
Now take the limit where the droplet spans a finite fraction of the whole
lattice; then its energy is predicted to grow as $N^{\theta/3}$ in three
dimensions if $N$ is the total number of spins. This is to be contrasted
with what happens in the mean field picture. In analogy with what occurs in
the Sherrington-Kirkpatrick model, the mean field picture suggests that
there are system-size excitations whose energy does {\it not} grow with $N$,
so effectively $\theta = 0$.

Now apply an {\it infinitesimal} field $B$. The magnetic term in the energy
can overcome the interfacial energy of a droplet of volume $s$ and force it
to flip. This will occur when the magnetic energy, of order $s^{1/2}B$,
becomes larger that the interfacial energy of order $s^{\theta/3}$. From
such a scaling, it is apparent that system size droplets are the most
unstable and flip for field intensities
\begin{equation}
B_{min} = N^{\theta/3-1/2}.
\label{eq_bmin}
\end{equation}
(In the case of the mean field picture, the same reasoning applies, but with
$\theta = 0$.) Both pictures thus predict the appearance of system size
events ($s=O(N)$) in the spectrum near $B = 0$, albeit at different scales
in $B$. Since $\theta$ is small, ($\theta \approx 0.2$~\cite{BrayMoore86})
the predictions for $B_{min}$ in the two different pictures are difficult to
distinguish numerically.

Interestingly, the situation is much more clear-cut when $B$ is finite
rather than infinitesimal because the two pictures then differ dramatically.
In that regime of magnetic field, there is a non-zero mean magnetization per
site which is proportional to $B$. In the droplet picture, this
magnetization makes it unfavorable to flip large scale clusters because the
magnetic energy term now grows as $sB^2$ and is of the wrong sign. Thus in
this approach, the sizes $s$ appearing in the spectrum are small (finite) as
soon as $B$ is non-zero. The field has introduced a length scale beyond
which there are no low energy excitations; the system is then paramagnetic.
Thus the spin glass phase does not survive when $B \ne 0$, and there is no
AT line. On the contrary, in the mean field picture, the energy landscape
continues to have many coexisting valleys even at finite values of $B$.
Furthermore, in this picture, these valleys have random and rather
independent energies at their bottoms. Then changing the magnetic field
should lead to level crossing of these valleys, so that the ground state is
expected to jump chaotically from one valley to another at high frequency,
generating events with $s=O(N)$ even when there is a mean magnetization.

\paragraph*{Finite size scaling ---}
Let us see what happens in reality. Our approach is based on the statistical
analysis of our data taken in the ensemble where the $J_{ij}$'s are Gaussian
random variables. Consider an $L \times L \times L$ lattice of $N$ spins and
a small interval $\left[ B, B+\Delta B\right]$. In the limit $\Delta B \to
0$, the frequency with which the ground states change in this interval is
proportional to $\Delta B$. Since the two pictures just discussed differ in
their predictions for events satisfying $s=O(N)$, we focus on the ``rate''
$r(N,B,x)$ (per $\Delta B$ and per $\Delta x$) at which a ground state is
changed by flipping $s=xN$ spins. To determine $r(N,B,x)$, we find for each
instance the ground states at regularly spaced values of $B$. For each
interval, we find the set of spins which have been flipped and find the
corresponding clusters defined as the connected components of that set. We
then cumulate the sizes $s$ of the events in a histogram, and average over
$10^3$ to $10^5$ randomly generated instances for lattice sizes $L=3, 4, 5,
6$ and $8$. Then the estimator for $r(N,B,x)$ is obtained by summing this
histogram for $s\in\left[(x-\Delta x/2)N,(x+\Delta x/2)N\right]$ and
dividing this sum by $\Delta B\Delta x$. We take $\Delta B=0.1$, a small
enough value so that the juxtaposition of several events is not mistaken for
one large event; we also set $\Delta x=0.1$. In view of the qualitative
discussion of valleys given in the previous section the signature of the
existence of the AT line is the survival of $r(N,B,x)$ at finite values of
$x$ and $B$ as $N\to\infty$. More precisely, for any fixed $x>0$ and $B\ne
0$, the droplet model predicts $r(N,B,x)\to 0$ as $N\to\infty$, whereas the
mean field picture predicts $r(N,B,x)$ does not tend to zero as $N\to\infty$
(as long as $B$ is not too large).

\begin{figure}[t]
\begin{center}
\resizebox{0.9\linewidth}{!}{\includegraphics{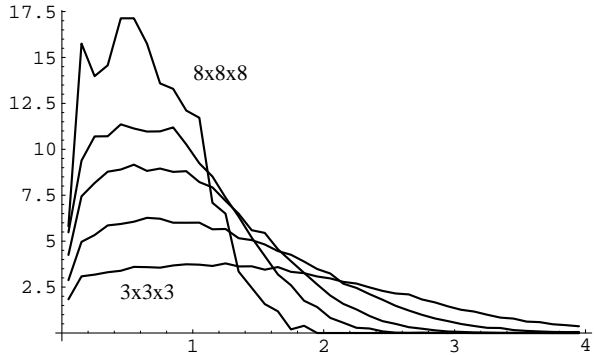}}
\end{center}
\caption{$r(N,B,0.15)$ as 
a function of $B$ for the EA model
with lattice sizes $3^3, 4^3, 5^3, 6^3$ and $8^3$.}
\label{fig_raw}
\end{figure}

\begin{figure}[t]
\begin{center}
\resizebox{0.9\linewidth}{!}{\includegraphics{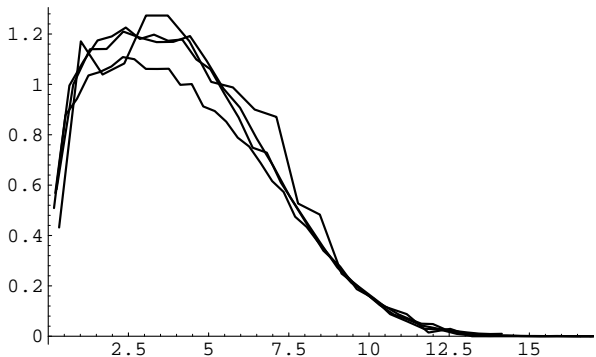}}
\end{center}
\caption{$r(N,B,0.15) / N^{\mu}$ as
a function of $BN^{\lambda}$ for the EA model with lattice sizes $4^3, 5^3,
6^3$ and $8^3$. We find $\mu=0.42$ and $\lambda=0.31$.}
\label{fig_fss}
\end{figure}

The data for $r(N,B,x)$ as a function of $B$ for different lattice sizes are
displayed in Figure~\ref{fig_raw} in the case $x=0.15$. Other values of $x$
give qualitatively similar results: the curves go to zero at large fields,
and as $N$ grows the curves cross at smaller and smaller values of $B$.
Furthermore it seems that for fixed $B$, $r(N,B,x)\to 0$ as $N\to\infty$. To
put such an extrapolation on a sound footing, we apply finite size
scaling, searching for a way to collapse all of the curves onto one-another.
For $N\ge 4^3$, we were able to do this by using a power law scaling:
\begin{equation}
r(N,B,x) \approx N^{\mu} R_x(BN^{\lambda}).
\label{eq_scaling}
\end{equation}
This is shown in Figure~\ref{fig_fss} for $x=0.15$, but the data collapse
occurs also for the other values of $x$, and with the {\it same} exponents.
The displayed scaling function falls to zero rapidly, seemingly faster than
any power, showing that large scale events become arbitrarily rare with
system size if the field is non-zero, contradicting the mean field picture.
These results indicate that the spin glass phase does {\it not} survive the
application of a finite magnetic field at zero temperature, in agreement
with the droplet picture prediction.

\paragraph*{Comparison to a mean field model ---}
The non-existence of a spin glass phase at $T=0$ and $B\ne 0$ is not
surprising from the point of view of the droplet picture. Were one to try to
defend the mean field picture of the spin glass phase persisting at finite
$B$, one would have to argue that the finite size scaling found in
Figure~\ref{fig_fss} is not perfect, and that larger system sizes might
begin to show that the correct interpretation of the data is different. Thus
it is useful to check that our proposed method of extrapolation is sensible
when considering a different model where we know that the mean field picture
is valid. To do so, we consider a dilute spin glass where there is no space
or geometry, but which is otherwise as close as possible to the
$3$-dimensional EA model. Thus we take the spins to be connected at random
while maintaining the connectivity fixed at $6$ for each spin. This
model~\cite{DominicisGoldschmidt89} is not very different from a Viana-Bray
model~\cite{VianaBray85}. The $J_{ij}$ couplings on the bonds which are
present are taken to have the same Gaussian distribution as in the lattice
model. The motivation for using this model rather than the Viana-Bray dilute
spin glass follows from the fact that the Euclidean (lattice) model and this
mean field model have the same Cayley tree approximation.

\begin{figure}[t]
\begin{center}
\resizebox{0.9\linewidth}{!}{\includegraphics{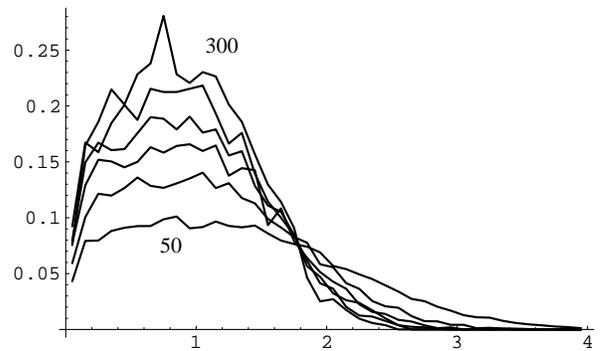}}
\end{center}
\caption{$r(N,B,0.15)$ as 
a function of $B$ for the mean field model 
with system sizes 50, 100, 250, 200, 250 and 300.}
\label{fig_mf_raw}
\end{figure}

We have analyzed this dilute spin glass model in the same way as we analyzed
the 3-dimensional lattice model. We find that the raw data shows a
significant difference with the Euclidean case: the different curves all
cross near the same point (see Figure~\ref{fig_mf_raw}). Performing finite
size scaling analysis, we find that the curves for different $N$ do not
superpose if we scale the field $B$, but do superpose on the left of the
crossing point if we use the scaling
\begin{equation}
r(N,B,x) \approx
\sqrt{N} Q_x(B),
\end{equation}
and this is true for all $x<0.5$.

\begin{figure}[t]
\begin{center}
\resizebox{0.9\linewidth}{!}{\includegraphics{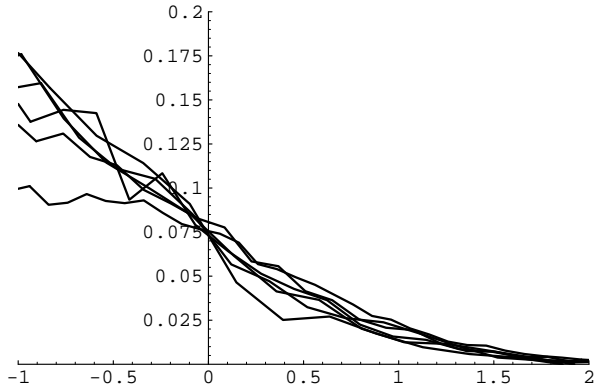}}
\end{center}
\caption{$r(N,B,0.15)$ as
a function of $\sqrt{N}(B-B^*_{0.15})$ for the mean field model
with system sizes 50, 100, 250, 200, 250 and 300. We find $B^*_{0.15}=1.79$.}
\label{fig_mf}
\end{figure}

The limiting large $N$ curve $Q_x$ falls to zero at a critical field $B^*_x$
which is identified by the crossing of the curves on
Figure~\ref{fig_mf_raw}. This is supported by a finite size scaling analysis
at large $N$: as shown in Figure~\ref{fig_mf}, the different curves
superpose when we plot $r(N,B,x)$ as a function of $\sqrt{N}(B-B^*_x)$. It
is intuitively clear and we find that $B^*_x$ decreases as $x$ grows; let
the largest value of $B^*_x$ be $B_c$ (we find $B_c\approx 2.1$), beyond
which there are no longer any events with $s=O(N)$. $B_c$ can be interpreted
as the crossing of the AT line with the zero temperature axis, so for
$B<B_c$ the system is in the spin glass phase and for $B>B_c$ it becomes
paramagnetic. (Note: in the Sherrington-Kirkpatrick model, the AT line goes
to $B=\infty$ as $T\to 0$. For our dilute spin glass $B_c(T)$ remains
bounded as $T\to 0$ because the model's connectivity is finite.)

\paragraph*{Discussion ---}
The picture to draw from our results is that in the presence of a finite
field, the 3-dimensional EA model behaves at zero temperature like a
paramagnet, as predicted by the droplet model. The spin glass phase is thus
destroyed by any finite magnetic field. This conclusion is in agreement with
recent experimental measurements~\cite{MattssonJonsson95}. To interpret this
result, it is natural to invoke the usual argument from the droplet model:
once the system has a mean magnetization $m>0$, it becomes nearly impossible
to find large scale objects to flip which have negative magnetization and
whose interfacial energy is not large. Nevertheless, we do not claim that
all of our results are explained by the droplet model. As discussed before,
the characteristic field intensity at which system size excitations arise is
given by Equation~\ref{eq_bmin}. Comparing with Equation~\ref{eq_scaling}
leads to $\lambda=1/2-\theta/3$. Since we find $\lambda\approx 0.31$, this
would correspond to $\theta\approx 0.57$ which is not compatible with the
commonly accepted value $\theta\approx 0.2$.

We have also investigated a dilute spin glass model which is the natural
mean field model for the 3-dimensional EA model. Although a finite magnetic
field gives rise to a finite magnetization, there is no associated length
scale, and the model maintains a spin glass phase up to some maximum field
intensity $B_c$. If we take at face value the droplet model interpretation
given above, we can say that since this mean field model has no space
embedding, large clusters are never compact, and because of that, it becomes
possible to find clusters of negative magnetization whose interfacial energy
is small. Our study of this model also shows that it is not the infinite
connectivity but rather the absence of geometry which allows the spin glass
phase to survive in the presence of a magnetic field.

\paragraph*{Acknowledgements ---}
We thank Jean Philippe Bouchaud for his very useful comments. J.~H.
acknowledges a fellowship from the MENESR, and O.C.M. acknowledges support
from the Institut Universitaire de France. The LPTMS is an Unit\'e de
Recherche de l'Universit\'e Paris~XI associ\'ee au CNRS.

\bibliographystyle{prsty}
\bibliography{/tmp_mnt/home/houdayer/Papers/Biblio/references}

\end{document}